\documentstyle[aps,preprint,tighten]{revtex}
\begin{document}
\draft
\preprint{\vbox{Submitted to {\it Physics Letters \bf{B}}\hfill
IFUSP/P-1240\\}}
%%%%%%%%% %%%%%%%%% %%%%%%%%% %%%%%%%%% %%%%%%%%% %%%%%%%%% %%
\title{\sc Virtual Meson Cloud of the Nucleon and Intrinsic Strangeness
and Charm}
\author { S. Paiva$^2$\thanks{e-mail: samya@if.usp.br}, 
\ M. Nielsen$^1$\thanks{e-mail: mnielsen@if.usp.br}, 
\ F.S. Navarra$^1$\thanks{e-mail: navarra@if.usp.br},
\ F.O. Dur\~aes$^1$\thanks{e-mail: dunga@if.usp.br}, \ and  
\ L.L. Barz$^1$\thanks{e-mail: lbarz@if.usp.br} \\ 
{\it $^1$Instituto de F\'{\i}sica, Universidade de S\~{a}o Paulo}\\
{\it C.P. 66318,  05389-970 S\~{a}o Paulo, SP, Brazil} \\[0.1cm]
{\it$^2$}Instituto de F\'{\i}sica Te\'orica, UNESP\\
{\it Rua Pamplona 145, 01405-901 S\~{a}o Paulo, SP, Brazil }}
\maketitle
\vspace{1cm}
\begin{abstract}
We have applied the Meson Cloud Model (MCM) to calculate the
charm and strange antiquark distribution in the nucleon. The resulting
distribution, in the case of charm, 
is very similar to the intrinsic charm momentum
distribution in the nucleon. This seems to corroborate the hypothesis 
that the intrinsic charm is in the cloud and,  at the same time,
explains why other calculations with the MCM involving strange
quark distributions fail in reproducing the low x region data. From the
intrinsic strange distribution in the nucleon we have extracted the
strangeness radius of the nucleon, which is in agreement with other
meson cloud calculations.
\\
PACS numbers 14.20.Dh~~12.40.-y~~14.65.-q
\\

\end{abstract}

\vspace{1cm}
%\newpage

Many years ago, it has been suggested by Sullivan \cite{sul}  that
some fraction of the nucleon's anti-quark sea distribution may be
associated with non-perturbative processes like the pion cloud of
the nucleon.  The generalization of this process to other mesons 
is depicted in figs. 1a and b, and was used in refs. \cite{thomas,koepf} to 
calculate the strange and anti-strange sea quark distributions in 
the nucleon.

Recent analysis of deep inelastic neutrino-hadron scattering data
\cite{ccfr} renewed the interest on the meson cloud picture of the
nucleon. It is well known that, in this picture there is an
asymmetry between sea quark and anti-quark momentum distributions 
\cite{thomas}.
This happens because the quark and the anti-quark are in different
hadronic bound states. On the other hand, in extracting sea 
distributions from hard processes, it is usually assumed that the
quark and antiquark sea contributions are equal. From the point
of view of QCD,  no definite statement on this subject can be made.
Based on charge conjugation symmetry it is only possible to say
that the quark sea distribution in the nucleon is equal to the
antiquark sea distribution in the antinucleon.
In ref. \cite{ccfr} it is shown that, in
contrast to the meson cloud approach expectation, the sea strange
and anti-strange quark distributions are quite similar. At first
sight this would be a very strong argument against the relevance
of the meson cloud \cite{ji}. The attempt to explain experimental
data with the meson cloud model performed in refs. \cite{thomas,koepf}
has shown not only that the asymmetry present in this model seems
to be in conflict with data but also that the calculated distributions
are far below data for $ x < 0.3 $. However, in ref. \cite{ma}, 
these data were
reconsidered and combined with the CTEQ collaboration analysis
\cite{cteq}. The conclusion of the authors was that, considering
the error bars, existing data do not exclude some asymmetry
between the strange and anti-strange momentum distributions, which is
significant only for $ x > 0.2 - 0.3 $.

In the present work we apply the meson cloud model (MCM) to study
strangeness and charm in the nucleon. In the case of strangeness, we shall
try to extract some estimates on the strangeness radius of the nucleon.
%From the experimental point of view 
This is a very interesting quantity from both theoretical and 
phenomenological point of view.
Indeed, approved  parity-violating lepton scattering
experiments at MIT-Bates \cite{mit} and CEBAF \cite{cebaf} will provide
information on the strangeness form factors at low and intermediate
$Q^2$ values, and more precisely determine the strange radius and
anomalous magnetic moment of the nucleon. In the antecipation of
the forthcoming data, considerable discussion about the strange matrix
elements of the nucleon, based primarily in nucleon models, have appeared 
in the literature. These nucleon model estimates 
\cite{ja,mus,cfn,hong,fnjc,koepf2,ham,hil,mi} contain large theoretical
uncertainties and their results may greatly differ  from each other.
For the strangeness (Sachs) radius, for instance, the predictions vary by 
over an order of magnitude and in their sign. A comparison of some of these
estimates can be found in ref.\cite{fnjc}.

We start drawing special attention to one point which, we believe, was
not enough stressed in ref. \cite{koepf,ma}, namely that 
there are  two kinds of $q \overline q$ fluctuations contributing
to the nucleon wave function: intrinsic and extrinsic. 
We shall emphazise, in particular, that we can identify the
distribution of the valence anti-quark in the meson cloud with the
distribution of the corresponding intrinsic anti-quark  \cite{brod}. 
Therefore, in order to fit data, one has necessarily to include the 
contribution of the extrinsic anti-quarks. 

In order to show the close relation between the meson cloud and 
the intrinsic
states we start in the charm sector where, due to large charm quark
mass, there is a striking difference between intrinsic and extrinsic
quark distributions. This makes
interesting the application of the MCM to charm. Whereas the case of 
strangeness we can confront MCM predictions directly with data, in
the case of charm we may confront a part of the MCM with the intrinsic charm
hypothesis (the eventual existence of accurate data on charm and 
anticharm distributions in the nucleon would allow us to test once 
more the MCM). 

We shall calculate the intrinsic nucleon $\overline c$ distribution,
${\overline c}_N^{(i)} (x)$, which, in the meson cloud approach is the
convolution of the valence $\overline c$ momentum distribution in
the ${\overline D}$ meson, ${\overline c}_D^{(v)} (x)$, with the 
momentum distribution of this meson in the nucleon $ f_{D} (y)$ :
\begin{equation}
x {\overline c}_N^{(i)} (x) = \int_{x}^{1}
dy\, f_{D} (y)\, \frac{x}{y} \, {\overline c}_D^{(v)} (\frac{x}{y})\; .  
\label{cn}
\end{equation}                       
The virtual ${\overline D}$ meson distribution in the nucleon's cloud
which characterizes its probability of carrying a fraction $y$ of the
nucleon's momentum in the infinite momentum frame is given by \cite{sul}
\begin{equation}
f_{D} (y) = \frac{g^2_{DN\Lambda}}{16 \pi^2} \, y \, 
\int_{-\infty}^{t_{max}}dt \, \frac{[-t+(m_{\Lambda}-m_N)^2]}{[t-m_D^2]^2}\,
F^2 (t)\; ,
\label{f}
\end{equation}
where $t=k^2$ is the meson virtuality. In the above equation $F(t)$ is a
form-factor at the $DN\Lambda$ vertex, and $t_{max}$ is the maximum value
of $k^2$, determined purely by kinematics:
\begin{equation}
t_{max} = m^2_N y- \frac{m^2_{\Lambda} y}{1-y}\; .
\label{tmax}
\end{equation}

For the $DN\Lambda$ form factor, following a phenomenological
approach, we use the monopole form:
\begin{eqnarray}
F (t) &=& \frac{\Lambda^2-m^2_D}{\Lambda^2-t}\; , 
\label{fmnb}
\end{eqnarray}
where $m_N$, $m_{\Lambda}$ and $m_D$ are the nucleon mass, the mass of the 
intermediate $\Lambda_c$ and the ${\overline D}$ meson mass respectively. 
$\Lambda$ is the form factor cut-off parameter. With this
parametrization we can directly compare the results obtained here
with those quoted in ref. \cite{nnnt}.

In the heavy quark effective theory \cite{man} it is assumed that the
heavy quark interacting with light constituents, inside a hadronic
bound state, exchanges momenta much smaller than its mass. Therefore
, to a good approximation, the heavy quark moves with the velocity
of the charmed hadron. Being almost on-shell, the heavy quark carries
almost the whole momentum of the hadron. These considerations suggest
that the $\overline c$ distribution in the
$\overline D$ meson is expected to be quite hard.
We shall start assuming that it is a delta function:
\begin{eqnarray}
{\overline c}_D^{(v)} (\frac{x}{y}) &\simeq& \delta\,(\frac{x}{y}-1) 
\,= \, x\, \delta\,(x-y)\; .
\label{del}
\end{eqnarray}
Evaluating (\ref{cn}) with the previously \cite{nnnt} used values
of $g_{DN\Lambda} =-3.795 $ and $\Lambda=1.2 $ GeV  we obtain for 
${\overline c}_N^{(i)}\,(x)$
the solid curve  shown in figure 2. For comparison we show with a
dashed line the intrinsic anti-charm distribution obtained by Brodsky
\cite{brod}:
\begin{eqnarray}
{\overline c}_N^{(i)} (x) &=& N x^2[\frac{1}{3}(1-x)(1 + 10x + x^2)
- 2 x (1+x) \ln \frac{1}{x}] \; ,
\label{intrin}
\end{eqnarray}
where $N$ is the same normalization constant used in \cite{brod}.
As it can be seen, there is a striking similarity between
both curves. Apart from details this result confirms our idea, that
the valence charm quarks in the cloud are intrinsic to the nucleon. 
It is also interesting to notice that with the parameters used above,
the total area below the curve in fig.~2, 
which gives the total intrinsic charm of the
nucleon, is $\sim 0.02$. This means that the two scenarios, the one
described in ref.\cite{nnnt} which evaluate the intrinsic charm by
relating it with the charm radius of the nucleon, and the one used here,
which extract the intrinsic charm directly from the $\overline c$
distribution of the nucleon, are consistent. They are also consistent
with the intrinsic charm hypothesis of ref.\cite{brod}.

In the strange sector, if we want the total $\bar{s}$ distribution in the 
nucleon we have to consider not only the valence $\bar{s}$ in the kaon, but
also the sea $\bar{s}$ quarks which are present in all hadronic intermediate
states. Therefore, we have to consider processes in the type of fig. 1b,
where the virtual photon strikes the recoiling baryon rather than the meson,
in addition to the processes in fig. 1a.
In ref.\cite{thomas} only the valence $s$ and $\bar{s}$ quarks in the
$\Lambda$ and $K$ intermediate states were considered when estimating
respectively the $s$ and $\bar{s}$ distributions in the nucleon. As argued 
in the previous part of this letter, this gives only the intrinsic
strangeness of the nucleon, and will fail to reproduce the total strange
distribution of the nucleon at small $x$ \cite{thomas}, where the extrinsic
part is important. Also in ref.\cite{thomas} it was assumed that the
amplitude to find an intermediate baryon with momentum fraction $y$ in
the nucleon (process of fig.1b) is the same as to find an intermediate meson
with momentum fraction $(1-y)$ in the nucleon (process of fig.1a). This
assumption has enhanced the asymmetry between $s$ and 
$\bar{s}$ distributions in ref.\cite{thomas}. The authors of ref.\cite{brown},
on the other hand, have neglected the off-shellness of the recoiling
baryon and have generated the baryon distribution in the nucleon's cloud.
Also, an additional parameter was introduced to ensure that the model 
generates an equal amount of strangeness and antistrangeness and, therefore, 
no asymmetry was produced. In ref.\cite{brown} only valence quarks were
considered in the intermediate states. In both works
\cite{thomas,brown} one expected to determine the sea quark distribution
in the nucleon from the knowledge only of the valence quark distributions of
the various intermediate states.

In ref.\cite{koepf} the authors
did consider the sea $\bar{s}$ (besides the valence $\bar{s}$)
distribution in the mesons, when calculating the $\bar{s}$ distribution
of the nucleon. However, they did not include the sea $\bar{s}$
distribution in the intermediate baryons, and this cannot be justified.
Of course the idea of
considering the sea quarks in the intermediate baryons to evaluate the
sea quark distribution in the nucleon leads to a self-consistent
equation since one of the possible intermediate states is just the
$N\pi$ pair. Besides, one has also to know the sea $\bar s$ distribution
in $\Delta,\;\Lambda,\;\Sigma,\;\Sigma^*$ and all other intermediate
baryons considered in the calculation, to extract the sea $\bar s$
distribution in the nucleon. Since there is no experimental knowledge
about all of them, it would be necessary to make further assumptions which
would reduce (or maybe destroy) the predictive power of the MCM.

So far our preliminary conclusion is that the MCM has a simple version
and may have a more complicated version. In the simple version (used
in the past by many authors \cite{thomas,brown,kumano}) we identify the
nucleon anti-quark content with the valence anti-quark in the meson
cloud of the nucleon. This simple hypothesis has been shown to work
well in the description of the large $x$ ($x>0.3$) domain of the anti-strange
distribution. The more elaborated version of the MCM should include the sea 
anti-quark component of the mesons and baryons. We believe
that with a proper inclusion of the sea, the MCM would be able to
account for the low $x$ data as well. However, according to the discussion
in the last paragraph, it is rather cumbersome to pursue this exercise, at
least
for the time being. Instead, we shall explore the simple version of the MCM.

In the first part of this work, studying charm distributions in the nucleon,
we have established the equivalence between the anti-quark distribution
obtained in the simple MCM with the intrinsic anti-quark distribution.
Now we make a further connection between the intrinsic strange content
of the nucleon and the strange squared radius.

In ref.\cite{nnnt} it was suggested that the probability of observing
the intrinsic Fock state $|uudq\bar{q}>$ is given by
\begin{equation}
P_{iq}={|r_q^2|\over|r_p^2|}\; ,
\label{piq}
\end{equation}
where $q=c,s$ or combined light flavors, $r_p$ is the average barionic
radius of the proton, $r_p=[<r_p^2>]^{1/2}\simeq0.72$ fm, and $r_q^2$ is
the average squared radius of the quark of flavor $q$. 

The ratio between the average squared radii in the above formula has
a simple geometrical interpretation. It gives us the relation between
the typical size of the fluctuations of a certain flavor $q$ and the
``total size'' of the nucleon, i.e., the size that takes into
account all possible fluctuations that couple to isoscalar currents.
Therefore, eq.(\ref{piq}) is a good measure of the probability that such
fluctuation occurs. Since the intrinsic
quark distribution functions in the nucleon give the probability to find 
the intrinsic quarks in the nucleon as a function of the fraction $x$ of 
the nucleon's momentum carried by the quark, 
% we can write
%\begin{equation}
%P_{iq}=\int_0^1 dx \bar q_N^{(i)}(x)\; .
%\label{pin}
%\end{equation} 
we can use eq.(\ref{piq}) to extract the strangeness
radius of the nucleon, by integrating its intrinsic strangeness distribution.

To evaluate the intrinsic strangeness of the nucleon we follow 
ref.\cite{koepf} and include only the
kaon as meson intermediate states, and the 
$\Lambda$, $\Sigma$ and $\Sigma^*$ as hyperon intermediate states. 
Therefore, we can rewrite Eq.(\ref{cn}) as
\begin{equation}
x {\overline s}_N^{(i)} (x,Q^2) =\sum_{Y} \int_{x}^{1}
dy\, f_{KY} (y)\, \frac{x}{y} \, {\overline s}_K^{(v)}(\frac{x}{y},Q^2)\; .  
\label{sn}
\end{equation} 

Since we have included decuplet states as hyperon intermediate states,
we need to generalize Eq.(\ref{f}) to account for this possibility.
Therefore, we write the kaon's distribution in the nucleon's cloud
as
\begin{equation}
f_{KY} (y) = \tau_Y\frac{g^2_{KNY}}{16 \pi^2} \, y \, 
\int_{-\infty}^{t_{max}}dt \, \frac{T(t,m_N,m_Y)}{[t-m_K^2]^2}\,
F^2_{KNY} (t)\; ,
\label{fmb}
\end{equation}
with
\begin{equation}
T(t,m_N,m_Y)= \left\{
\begin{array}{cc} 
-t+(m_Y-m_N)^2 & Y\in 8
\\
{((m_Y+m_N)^2-t)^2((m_Y-m_N)^2-t)\over 12m_N^2m_Y^2} &  Y\in 10
\end{array}  \right.\; ,
\label{t}   
\end{equation}
for an intermediate octet or decuplet hyperon. In the above equation  
$t_{max}$ is still given by Eq.(\ref{tmax}) only with $m_{\Lambda}$
substituted by $m_Y$, the intermediate hyperon mass. In Eq.(\ref{fmb}),
$\tau_Y$ is an spin-flavor SU(6) Clebsh-Gordon factor. 
 For the kaon-hyperon couplings we follow ref.
\cite{koepf} and relate them to the pion-nucleon couplings using 
spin-flavor SU(6) \cite{holz}. The form factor, $F_{KNY}(t)$, at the 
nucleon-kaon-hyperon vertex is usually parametrized either in monopole,
dipole or exponential form. To make a direct comparison with previous
work easier, we will restrict ourselves to the exponential form:
\begin{equation}
F_{KNY} (t) = e^{(t-m_K^2)/\Lambda^2_{KNY}}\; ,
\label{form}
\end{equation}
with $\Lambda_{KNY}=1200$ MeV \cite{koepf}.

We neglect the small difference between the pion and kaon structure
functions and use the better determined pion structure function
to evaluate eq.(\ref{sn}).
In figure 3 we present the $\overline s_N^{(i)}$ distribution  
obtained with the help of the SMRS \cite{na10} parametrization of the pion 
structure function extracted from fits to Drell-Yan pair production
experiments. From this figure we get $P_{is}=0.12$. It is very interesting
to notice that the authors of ref.\cite{dg}, using the MIT bag model to 
calculate the probability of finding a five-quark component $|uudq\bar{q}>$
configuration bound within the nucelon bag, got $P_{ic}=0.02$ and 
$P_{is}=0.16$, in good agreement with our results.

Comparing fig.3 with fig.2 we find that the intrinsic charm carries more
momentum than
the intrinsic strangeness, as expected. The shape of the intrinsic 
strangeness distribution seems very similar to the usual parametrizations
of sea-quark distributions. However, as it can be seen in fig.4, it has a very
important difference with respect to the real sea distribution: 
$x{\bar s}^{in}_N(x)$
goes to zero as $x$ goes to zero. In fig.4 we also show the CCFR data
\cite{ccfr}, and the $\overline s$ distribution calculated in 
ref.\cite{koepf} with the help of
the GRV \cite{na24} parametrization for the pion structure function.
From this figure we clearly see that for $x>0.3$ the intrinsic distribution
can fairly well describe the data.

In fig.3 we also plot the individual
contribution of each intermediate pair considered in our calculation.
As can be seen, the contribution from the $\Sigma-K$ pair is much smaller
than the others, due to the smaller coupling constant. This was expected
and this contribution was even neglected in refs.\cite{mus,cfn,fnjc} due
to this reason. However, the contribution of the $\Sigma^*-K$ pair is
of the same order of magnitude as the $\Lambda-K$ pair and cannot be
neglected. Using the value $P_{is}=0.12$ in the above relation,
eq.(\ref{piq}), we get for the strangeness radius
\begin{equation}
|<r_s^2>|=6.22\times10^{-2} fm^{2}\; .
\end{equation} 
If we still add to it the vector-meson dominance model contribution 
coming from the $\omega-\phi$ mixing \cite{cfn} we get
\begin{equation}
|<r_s^2>|=7.97\times10^{-2} fm^{2}\; ,
\end{equation}
 which is about two times the result in ref.\cite{cfn} and only less than 
half of the result in ref.\cite{ja}. It is worth mentioning that we can not 
predict the sign of the square strangeness radius in our approach.

If we had considered only the $\Lambda-K$ loop in our calculations, instead
of  $\Lambda-K,\; \Sigma-K$ and $\Sigma^*-K$ as we did, we would have got
$P_{is}=0.0585$. This would reduce the strangeness radius to
$|<r_s^2>|=3.03\times10^{-2}$ fm$^{2}$, which is completely consistent
with the medium value obtained in ref.\cite{mus}: 
$|<r_s^2>|=2.5\times10^{-2}$ fm$^{2}$. This value was obtained by 
probing the $\Lambda-K$ loop with a vector strange current.  

To summarize, based in our reslts, we would like to stress that the 
strangeness radius
of the nucleon will be experimentally accessible in the near future. From
the theoretical point of view it is not yet clear how to calculate it
precisely. Some of the theoretical works concentrate on the computation
of the strange vector form factors from mesonic loops. So far a complete 
computation of all loops, including all the relevant hadrons, is not
available. Our estimates, based on the intrinsic strangeness of the nucleon,
are useful guides, and give us the relative importance of each mesonic loop.
In particular our calculation strongly indicates that the $\Sigma^*-K$
loop should be included in form-factor computations \cite{fmnn}.
                      
\vspace{0.5cm}

\underline{Acknowledgements}: This work has been supported by FAPESP
and CNPq.  We would like to warmly thank S.J. Brodsky and W. Koepf 
instructive discussions. 

\vspace{1cm}
\newpage
\noindent
{\bf Figure Captions}\\
\begin{itemize}

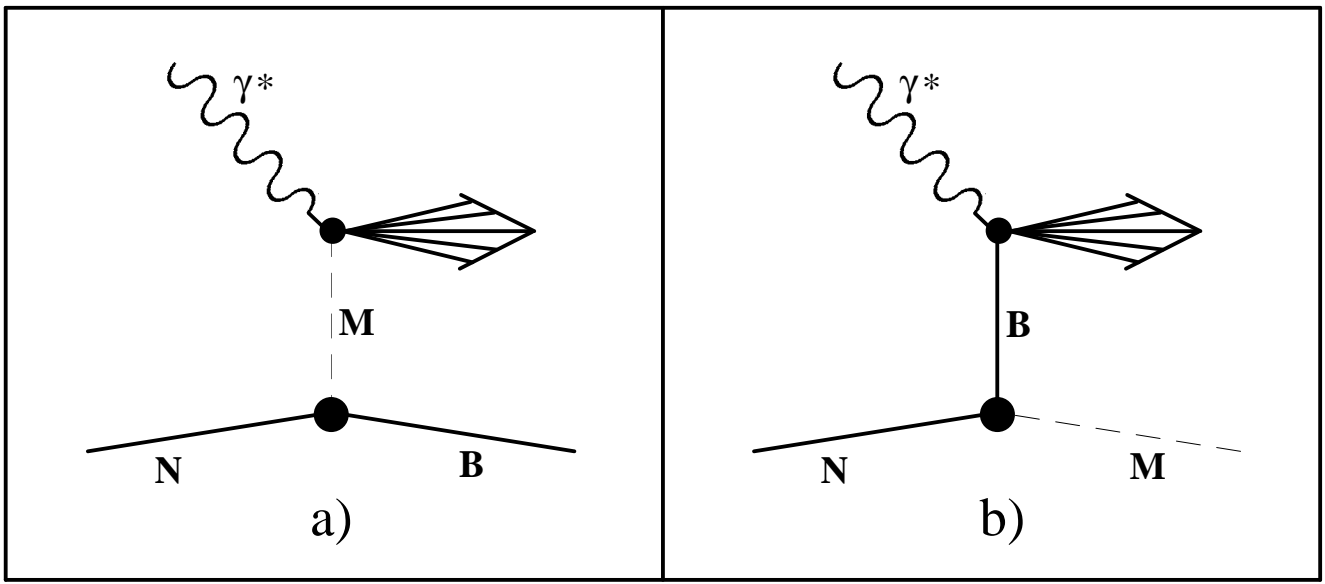
\item[{\bf Fig. 1}] The generalized Sullivan proccess. a) the virtual
photon strikes the cloud meson. b) the virtual photon strikes the
recoiling baryon.

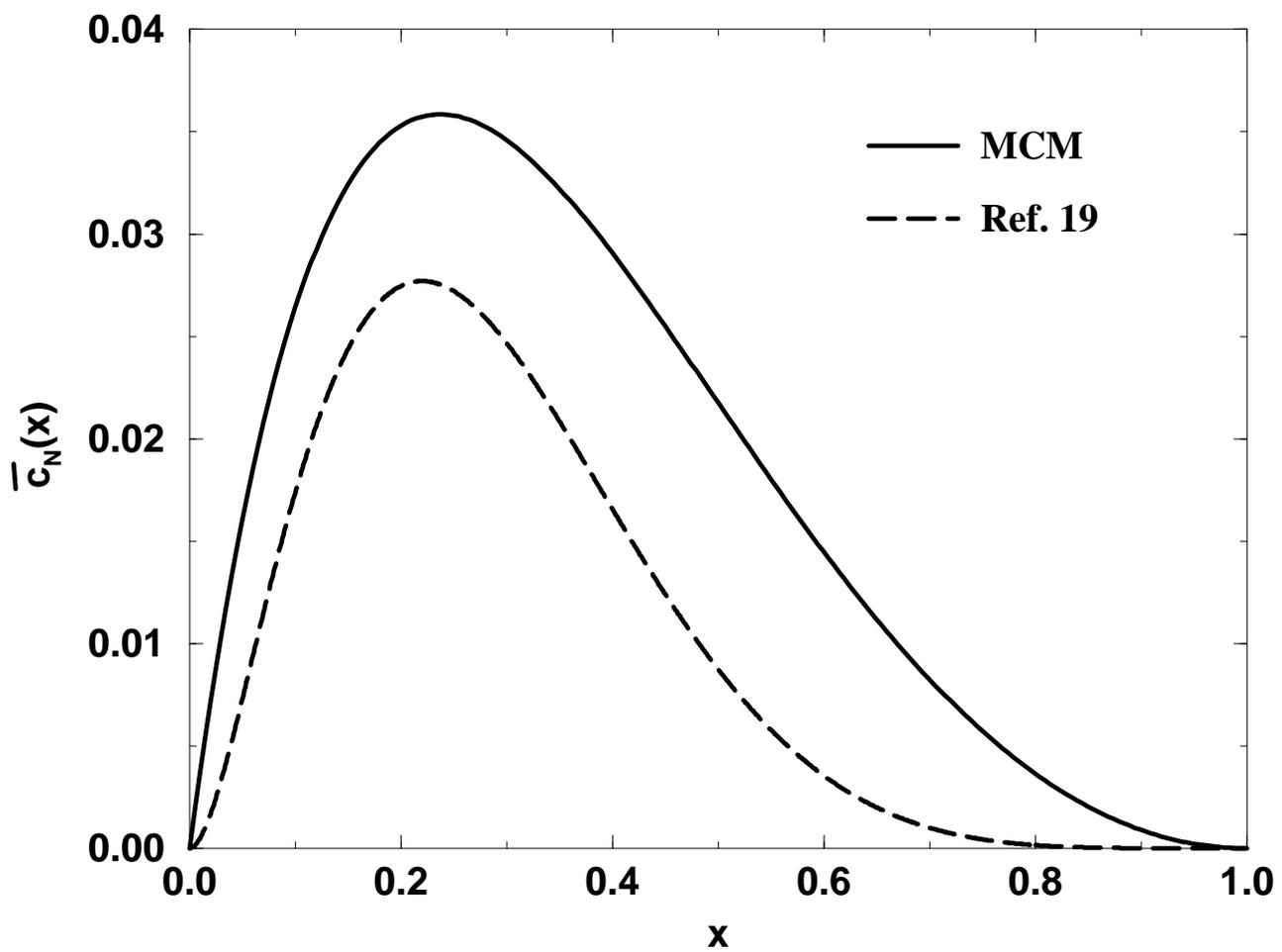
\item[{\bf Fig. 2}] ${\overline c}_N\,(x)$ distribution calculated
with the meson cloud model (solid line) and the intrinsic anti-charm 
distribution (dashed line).

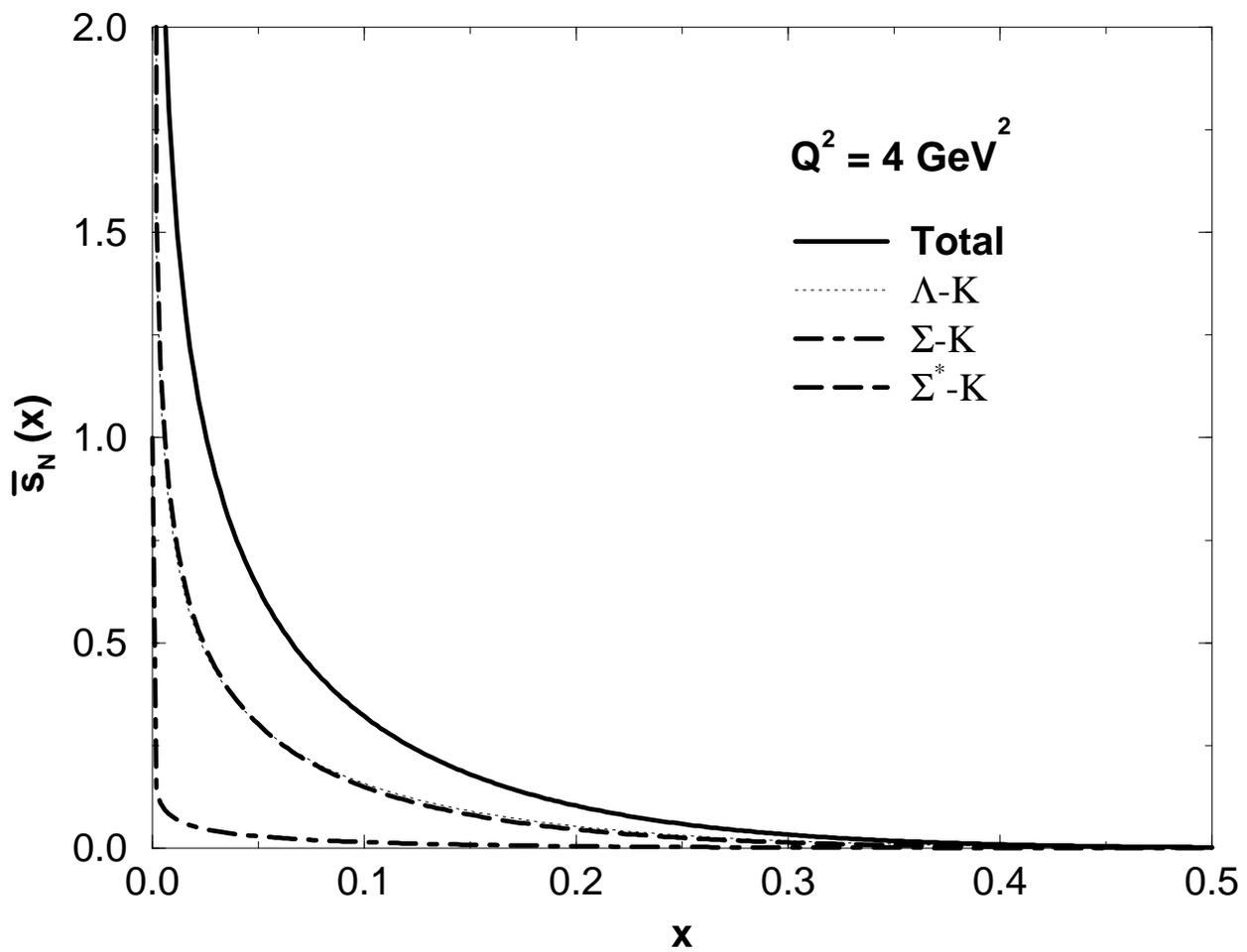
\item[{\bf Fig. 3}] Intrinsic ${\overline s}_N\,(x)$ distribution calculated
with the meson cloud model using the NA10 pion structure function
(solid line). The individual contributions
are: $\Lambda-K$ (dotted line), $\Sigma-K$ (dot-dashed line) and
$\Sigma^*-K$ (dashed line).

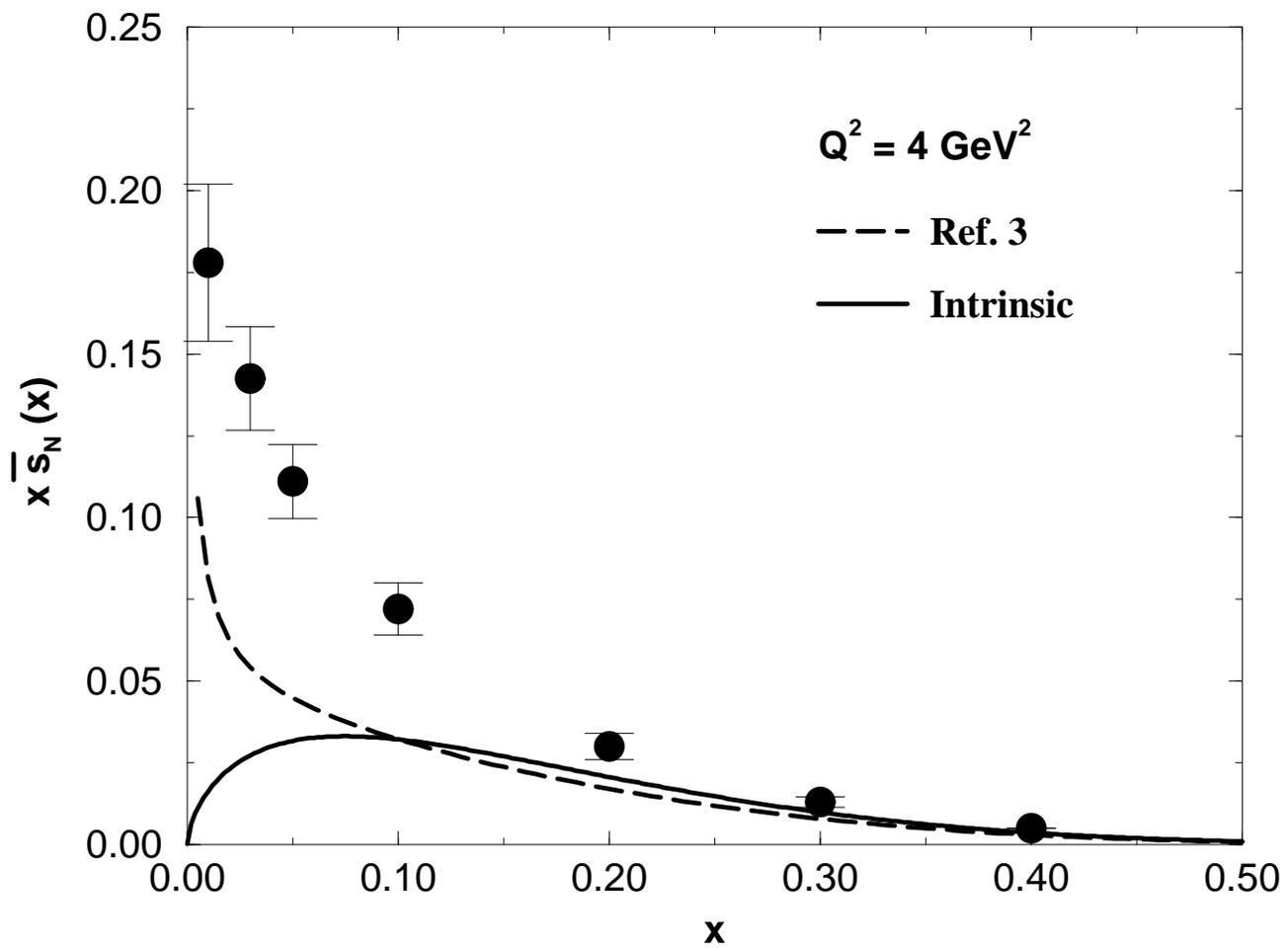
\item[{\bf Fig. 4}] Intrinsic $x{\overline s}_N\,(x)$ distribution calculated
with the meson cloud model using the NA10 pion structure function 
(solid line). The data are from the CCFR collaboration and the dashed line
shows the calculation of ref.[3].

\end{itemize}

\end{document}